\documentclass[10pt,journal,twoside]{IEEEtran}
\usepackage{graphicx}
\usepackage{float}
\usepackage{array}
\usepackage{amsmath}
\usepackage[export]{adjustbox}
\usepackage{caption}
\usepackage[justification=centering]{caption}
\usepackage{multirow}
\usepackage{multicol}
\usepackage{booktabs}
\usepackage{hhline}
\usepackage{xcolor}

\usepackage{mathtools}

\DeclarePairedDelimiter\abs{\lvert}{\rvert}%
\DeclarePairedDelimiter\norm{\lVert}{\rVert}%

\makeatletter
\let\oldabs\abs
\def\abs{\@ifstar{\oldabs}{\oldabs*}}
\let\oldnorm\norm
\def\norm{\@ifstar{\oldnorm}{\oldnorm*}}
\makeatother

\ifCLASSOPTIONcompsoc
  \usepackage[nocompress]{cite}
\else
  \usepackage{cite}
\fi

\begin{document}
 \pagenumbering{gobble} 
\title{Energy-Efficient Real-Time Heart Monitoring on Edge-Fog-Cloud Internet-of-Medical-Things}

%Real-Time Quality-Aware Energy-Efficient Heart Monitoring on an Internet-of-Medical-Things}
\author{Berken Utku Demirel, Islam Abdelsalam Bayoumy,  Mohammad Abdullah Al Faruque, \IEEEmembership{Senior Member, IEEE}
	\thanks{The authors are with the Department of Electrical Engineering and Computer Science, The Henry Samueli School of Engineering, University of
			California at Irvine, Irvine, CA 92697-2625 USA (e-mail: bdemirel@uci.edu).}}

% make the title area
\maketitle

\begin{abstract}
The recent developments in wearable devices and the Internet of Medical Things (IoMT) allow real-time monitoring and recording of electrocardiogram (ECG) signals. However, continuous monitoring of ECG signals is challenging in low-power wearable devices due to energy and memory constraints. Therefore, in this paper, we present a novel and energy-efficient methodology for continuously monitoring the heart for low-power wearable devices. The proposed methodology is composed of three different layers: 1) a Noise/Artifact detection layer to grade the quality of the ECG signals; 2) a Normal/Abnormal beat classification layer to detect the anomalies in the ECG signals, and 3) an Abnormal beat classification layer to detect diseases from ECG signals. Moreover,  a distributed multi-output Convolutional Neural Network (CNN) architecture is used to decrease the energy consumption and latency between the edge-fog/cloud. Our methodology reaches an accuracy of 99.2\% on the well known MIT-BIH Arrhythmia dataset. Evaluation on real hardware shows that our methodology is suitable for devices having a minimum RAM of 32KB. Moreover, the proposed methodology achieves  $7\times$ more energy efficiency compared to state-of-the-art works. 
\end{abstract}


\section{Introduction}

\IEEEPARstart{E}{lectrocardiogram (ECG)} signals are widely used to detect cardiovascular diseases, which are the leading cause of death globally \cite{Lancet}. Moreover, according to the American Heart Association, the early detection of these diseases is crucial for patients' health \cite{AHA}. Clinical ECG is the primary tool for monitoring cardiac activity. However, it can only be used for a limited time, and continuous monitoring of the patients' condition is still required outside clinical hours. Traditionally, ambulatory ECG devices are used to monitor the cardiac activity for a long duration to be further investigated by clinicians. For example, the Holter \cite{Holter}, a battery-operated portable device, is used to record and store long-term ECG signals. However, these devices cannot provide real-time feedback to users, and cardiologists need to analyze long-term recordings, which is a very time-consuming and expensive process.
\\
To solve this issue, several heart monitoring devices and solutions have been proposed and developed both in academia and industry \cite{Nature2}, \cite{IoMT}, thanks to rapid development on the internet of Medical Things (IoMT) and smart health care systems. These heart monitoring devices or systems can be categorized into two different groups according to their methods. The first group \cite{Nature},\cite{Remote-2} analyzes long-term recorded ECG signals offline by using remote cloud servers. The utilized algorithms in the cloud provide a powerful classification performance. However, they cannot be implemented on the edge node due to their memory requirements and high energy consumption.
Moreover, since all computing occurs in the cloud, the latency of the system increases, which weakens the user experience \cite{IoT-Latency}. The latency is an essential factor for heart monitoring applications because the rapid detection of cardiovascular diseases is critical for people's lives.
%\\ 
The second group \cite{Real-time1,Real-time2} provides a real-time solution by doing computation on the edge side rather than the cloud; however, the amount of time the device is monitoring the cardiac activity is limited due to constraints on battery life, which is the most valuable resource for the edge of the network \cite{IoT-Battery}.
\\
Due to these problems, many people live for years unaware of their illness \cite{AHA2}. Some cases reported that deaths due to cardiovascular disease could have been prevented if the disease was detected earlier \cite{preventable}. Therefore, continuous real-time ECG monitoring can be a vital solution for people with cardiovascular diseases.
\\
The main challenge of designing a continuous real-time monitoring system is adhering to the devices' energy and memory constraints since processing requires lots of memory and is computationally intensive. To overcome this problem, existing works have proposed performing the computation at the proximity of data generation sources, which are the edge devices in this case, using fog or edge-cloud \cite{Edge-Cloud,Edge-Cloud2,Nafiul} architectures by transferring real time signals to these nodes of network. However, this transfer operation requires tremendous communication power, which decreases the device's battery life and makes it difficult to sustain continuous monitoring over long periods. This kind of IoMT system requires significant energy resources on the edge device, and is vulnerable to privacy issues\cite{IoT-Security,IoT-Security2}. Several methods have been proposed to encrypt the ECG signals. However, they require additional  energy and memory at the edge  \cite{Compression}, \cite{Compression2}.
Moreover, the transferred ECG signals might be contaminated with noise or artifacts caused by the users' mobility, which results in further unnecessary energy consumption. In summary, the key research challenges associated with continuous heart monitoring are:  (i) {\it developing resource-efficient algorithms on the edge}, (ii) {\it detecting abnormalities in ECG signals as fast as possible to minimize latency}.

To address the above mentioned challenges, this paper proposes a novel heart monitoring methodology on a hybrid edge-fog/cloud IoMT. In this paper, we define ``edge''  as the computing platform where the data acquisition is performed, and ``fog'' is defined as the device possessing further computing and network resources along the path between data sources and cloud data centers, e.g., a smartphone. Throughout the paper, ''edge-fog/cloud'' is used since the proposed methodology is a solution for any two-tiered systems employing edge-cloud, edge-fog, or fog-cloud architectures. The proposed methodology delivers a layered software pipeline architecture by distributing the layers between edge and fog/cloud. The first layer running on the edge is designed to detect Noise/Motion artifacts. This detection aims to conserve energy resources through avoiding unnecessary artifacts transmission. The second layer, running as well on edge, classifies normal and abnormal beats in the ECG. If the beats are classified as normal, the classification is considered complete, and only the heart rate value of that beat is transmitted instead of the raw signal. Else, the beats are classified as abnormal and are sent to the next layer in the hierarchy, which could be a fog or cloud, to be further classified.
Moreover, these first two-layers can further reduce the energy consumption of the edge device through controlling the data sampling rate. If the signal is not classified as clean or the recorded ECG signal has no abnormal beats, the control unit changes the sampling rate to the degree that both classifiers still maintain high-performance classification. The last layer, running on fog or cloud, is a distributed multi-output CNN to classify several cardiovascular diseases.

\vspace{-3mm}
\subsection{Motivational Example}
\begin{figure}[h]
    \centering
    \includegraphics[scale = 0.4]{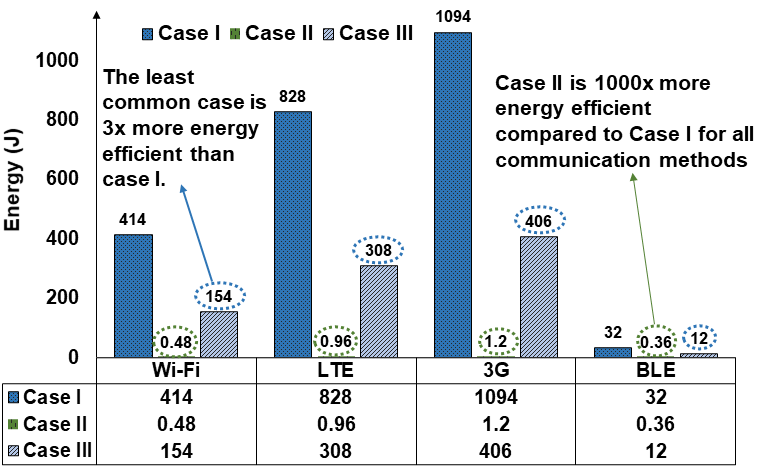}
    \caption{Energy consumption of Case I, II, and III}
    \label{fig:Comm_energy}
\end{figure}

We have done several experiments to show the advantages of the proposed methodology for continuous heart monitoring. In  Case  I,  the  1-hour  raw  ECG, which are acquired digital signals from sensors,  are transferred to the fog/cloud server without any investigation or operation on the edge for noise and artefacts. Then, the energy consumption of these transfer operations is calculated for four different communication technologies, \textit{Wi-Fi}, \textit{LTE}, \textit{3G} and \textit{BLE}. For Case II, the 1-hour ECG record is partitioned  into  two  parts:  i)  a  50-minutes  segment  of  clean ECG which has no artifacts with  just  regular  beats and ii) a 10-minutes segment of the ECG signal, simulated as a noisy signal. The required communication energy for transferring this record is calculated again. Lastly, in Case III the ECG signal is divided into three parts: i) a 40 minutes regular beats segment; ii)  a 10 minutes noisy signal; and iii) 10 minutes of the recording including  several arrhythmias. The energy consumption for all cases' transfer operation is shown in Figure \ref{fig:Comm_energy}. While calculating the energy consumption of these cases, we have followed the $\mu J/bit$ values given in \cite{COMM} for the \textit{Wi-Fi}, \textit{LTE}, and \textit{3G}. For \textit{BLE} protocol energy consumption, we performed the profiling for data exchange on an EFR32BG13 Blue Gecko Bluetooth® Low
Energy SoC which has 32-bit ARM Cortex-M4 core with 40 MHz maximum operating frequency.

Since the raw ECG data is transmitted without any classification on edge in Case I, its communication energy is the highest among the three cases regardless of the communication methods as shown in Figure \ref{fig:Comm_energy}. However, when we applied our proposed methodology to detect the normal and noisy beats in the ECG signal and filtered out them before transmitting to the fog/cloud node (Case II, III), the communication power can be decreased by $1000\times$ and $3\times$ for Case II and III, respectively. Also, to make a fair comparison with different cases, we measure the additional computational energy consumption of our proposed methodology to detect the artifacts and abnormal beats in the ECG signals for case II and III. We observe that the algorithm's energy consumption approximately 1.5 J for an hour, which makes Case II and III still much more energy-efficient than Case I. Moreover, it is known that arrhythmias are not as frequent in people like Case III (ten minutes of hour), so case II will be more common for most of the people. Therefore, we observe that detecting the normal and noisy ECG beats in edge devices could save energy and time while reducing the communication channel usage and fog/cloud server load.
\vspace{-2mm}
\subsection{Novel Contributions}
The novel contributions of this paper are as follows:
\begin{itemize}
\item A real-time continuous heart monitoring system that runs on a hybrid edge-fog/cloud architecture while being energy and memory efficient.
\item A novel layer-wise distributed multi-output CNN architecture that is optimized for decreasing the energy consumption and latency between edge and fog/cloud. To the best of our knowledge, we are the first to investigate the signal's physiology in distributing the computational complexity of architecture between the different nodes of a network.
\item Evaluation of our proposed methodology on the well known MIT-BIH datasets \cite{Physionet-MIT-BIH}, and PhysioNet/CinC (PICC)  2011  challenge  \cite{Physionet-SQA, Physionet}. It shows that, our proposed methodology reaches or outperforms the current state-of-the-art works in terms of classification performance\cite{EMBC,Access, BCS, Behar, IoT-SQA}.
\item Evaluation on real hardware shows that each layer of the proposed methodology achieves up to $7\times$ more energy efficiency compared to the state-of-the-art works \cite{EMBC, Access, BCS}.

\end{itemize}

The rest of the paper is organized as follows. In Section \ref{related-work}, we review related works for heart monitoring systems. Section \ref{method} describes the proposed methodology. Section \ref{setup} shows the experimental setup. Section \ref{result} discusses the results. Finally, a conclusion is drawn in Section \ref{conclusion}.

\section{Related Work}
\label{related-work}
\subsection{Quality Assessment of ECG}

Signal Quality Assessment (SQA) is the critical first step in continuous heart monitoring since it eliminates the noisy signals before the classification. Mostly, SQA methods grade ECG into two groups: $acceptable$ and $unacceptable$.  Existing methods extracted several features from the ECG signals and graded them using heuristic rules \cite{IoT-SQA} or machine learning-based classifiers such as Support Vector Machine \cite{Behar}, \cite{IoT-SQA2}.  For example, authors in \cite{Behar} proposed extracting six different features based on time and frequency domains, whereas the extraction of these features depends on the accurate and reliable detection of the QRS complex in noisy ECG signals, which is a challenging task. To avoid detection of QRS complex,  authors in \cite{IoT-SQA} examined three main causes of noise in ECG: (i) abrupt change, (ii) signal absence, and (iii) high-frequency noise. Since the authors do not detect the QRS complexes in ECG signals, they investigated the 10-second windows based on time and frequency. However, since the number of samples increases with a longer duration of ECG recording, the feature extraction-related algorithms' computation requirements increase, especially when the frequency-domain features are used. Since these SQA algorithms run in the proximity of data sources (edge of the networks) during acquisition, they need to be energy and memory efficient. Therefore, in this paper, we proposed a lightweight SQA which can detect noise and artifact.

\subsection{Heartbeat Classification of ECG Signals}
A wide range of automatic heartbeat classification methods has been proposed, and these existing methods may be categorized into two groups. The first method \cite{Remote-2, Real-time1} extracts some handcrafted features from the ECG signals and feeds them to a classifier like SVM. For example, authors in \cite{Remote-2}, extract 6 frequency-domain features from the heart rate for arrhythmic beat classification. Similarly in \cite{Real-time1} authors use 32 time-domain features with linear kernel SVM to classify heartbeats. However, this feature extraction process increases the computational complexity and memory. Moreover, these extracted features may not represent the complete characteristics of the ECG and restrict the performance. 
\\
The second method is to directly send the original waveform to a neural network for classification, known as end-to-end classification method to avoid the feature extraction process since the neural networks do not require feature engineering as they automatically extract features.  For example, in \cite{EMBC}, authors use a combination of Bidirectional Recurrent Neural Network (BRNN) and CNN model to detect 4 kinds of heartbeats. Compared to feature-based classification methods, this method can reach higher accuracy. However, this combined architecture has tens of millions of parameters, making it unsuitable for the edge node with memory and energy constraints. Similarly, authors in \cite{BCS} proposed a two-stage neural network, which combines a multi-layer perceptron and a CNN. The first stage classifies the ECG beats as normal or abnormal, and the second stage classifies the abnormal beats to several arrhythmias. Moreover, their proposed solution uses an additional classifier for discriminating abnormal beats, which introduces more parameters and energy consumption to the system. However, our proposed methodology uses a distributed multi-output CNN architecture to detect and filter out the regular beats during run-time without using any additional features and classifiers. 

\section {Our Proposed Methodology}
\begin{figure*}[ht]
	\centering
	\vspace{-11mm} \includegraphics[scale = 0.5]{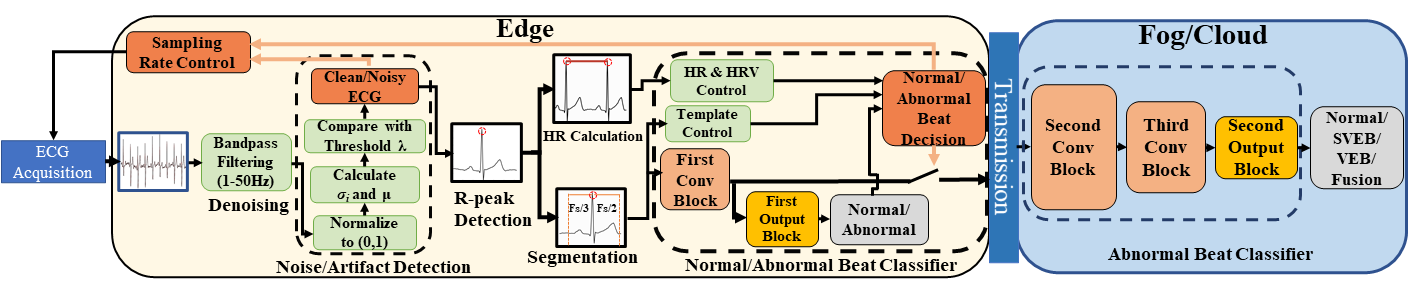}
	\caption{Overview of our proposed methodology}
	\label{Methodology}
\end{figure*}

The proposed methodology, shown in Figure \ref{Methodology}, consists of 3 main layers and several processing blocks whose components are detailed in the following section.
\label{method}

\subsection{Processing and Control Units}
\label{processing}
\subsubsection{Filtering}
We use a fifth order linear phase bandpass filter with a Hamming window cut-off frequencies ($f_1$=1 Hz and $f_2$=50 Hz).

\subsubsection{R-peak Detection}
To detect the R-peaks, we have used the Pan-Tompkins algorithm \cite{Pan-Tho}, a real-time QRS complex-based heartbeat detection approach that has an accuracy of up to 99.5\%.

\subsubsection{Beat Segmentation}
We take $F_s/3$ samples before and $F_s/2$ samples after the R-peak. The highest heart rate is chosen as 180 Beats Per Minute (BPM) to avoid overlapping two beats in a window.

\subsubsection {Heart Rate Calculation}
\label{hr_calculation}
The heart rate is calculated using the following equation:
\begin{equation}
\label{hr_eq}
HR_{(i)}=\frac{60}{(R_{(i)}-R_{(i-1)})/F_s}
\end{equation}
Where $HR_{(i)}$ is the heart rate of the $i^{th}$ heartbeat segment in BPM. $R_{(i)}$ is the location of the R-peak in the $i^{th}$ heartbeat segment, and $F_s$ is the sampling frequency.

\subsubsection{Sampling Rate Control}
This unit controls the ECG acquisition sampling rate according to the output of two different decision units.  The first decision unit checks the ECG signal's quality. If it detects a noisy ECG signal, the sampling rate is decreased to the degree that all components maintain the high classification performance. The second decision unit monitors the incoming beats and classify them as normal and abnormal beats. If the beats are classified as normal, this unit decreases the sampling rate to the same degree. 

\subsection{Noise/Artifact Detection Layer}
The ambulatory ECG signals are mostly contaminated with low-frequency motion artifacts that cannot be removed using simple filtering. Therefore, the first layer is designed to detect these artifacts. Since this algorithm runs on the edge, we have focused on developing a lightweight and robust algorithm to increase the device's battery life. We have observed that a rule-based decision method is superior compared to the classical machine learning algorithms (SVM, RF) considering energy efficiency while maintaining the classification performance for this task. Therefore,  a rule-based algorithm is used to classify signals into two groups $acceptable$ and $unacceptable$.
The ECG signals are divided into 10-seconds windows. First, the windows are normalized with respect to the maximum amplitude value. The mean of a normalized window is obtained and compared with a threshold $(\lambda)$ to detect the absence of an ECG signal. If the mean is lower than the threshold, the window is classified as $unacceptable$ and the signal is discarded.

% \vspace{-5mm}
% \begin{multicols}{2}

% \begin{equation*}
%     \mu = \frac{1}{N} \sum_{n=0}^{N-1}x[n] 
% \end{equation*}\break
% \begin{equation}
%      flag = \begin{cases}
%          0  \hspace{2mm} \lambda>\mu\\  
%         1  \hspace{2mm} \lambda \leq \mu
%           \end{cases}
%     \label{eqn:FL}
% \end{equation}
% \end{multicols}

The abrupt changes and baseline wander are investigated using a moving standard deviation of the 10-seconds ECG signal. In this method, a window of a specified length ($2F_s/5$) is moved over the signal with a 70\% overlap. The deviation of the signal ($\sigma_i$) is computed over the data by using Equation \ref{movstd}.

\begin{equation}
    \sigma_{i} = \sqrt{\frac{1}{N-1}\sum_{n=1}^{N}\lvert x_i[n] - \mu_i \rvert ^2}
    \label{movstd}
\end{equation}

\begin{figure}[h]
    \centering
    \includegraphics[width=9cm, height=8cm,inner]{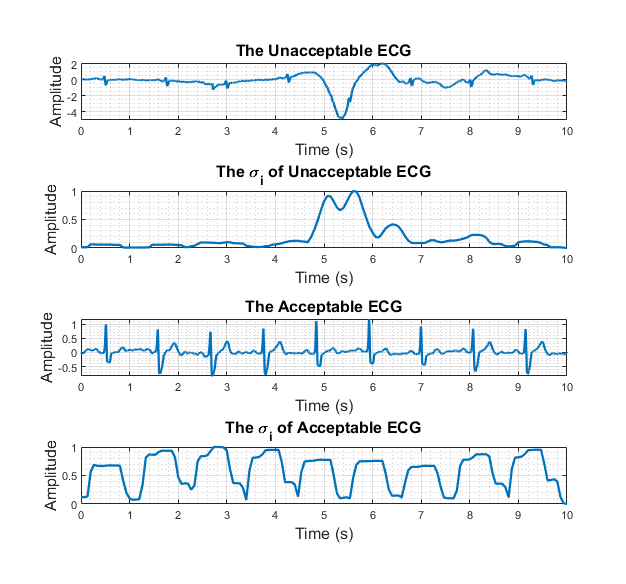}
    \vspace{-10mm}
    \caption{The $Acceptable$ and The $Unacceptable$ ECG signals from PICC \cite{Physionet-SQA} and obtained $\sigma_i$ waveforms}
    \label{fig:Both}
\end{figure}

Where $\sigma_i$ and $\mu_i$ are the corresponding standard deviation and mean of the window, respectively.
An example of $\sigma_i$ waveforms is shown in Figure \ref{fig:Both}. When the ECG signal has an abrupt change, the algorithm suppresses the beats and brings the artifacts forefront. Finally, to detect the abrupt changes, the mean of the waveform ($\sigma_i$) is obtained and compared with a threshold, which is set to $0.2$ based on acceptable level of noise. If the mean is higher than the threshold, the signal is classified as acceptable.
\vspace{-4mm}

The proposed method has the following advantages: (i) it is sensitive to abrupt changes and baseline wanders where if a small portion of the signal is corrupted with an artifact or noise, it is detected and ignored; (ii) it is energy efficient since it can run at a low sampling rate of 100-120 Hz without losing performance, and it does not need complex features to classify.

\subsection{Normal/Abnormal Beat Classification}
\begin{figure}[H]
    \centering
    \includegraphics[width=0.38\textwidth]{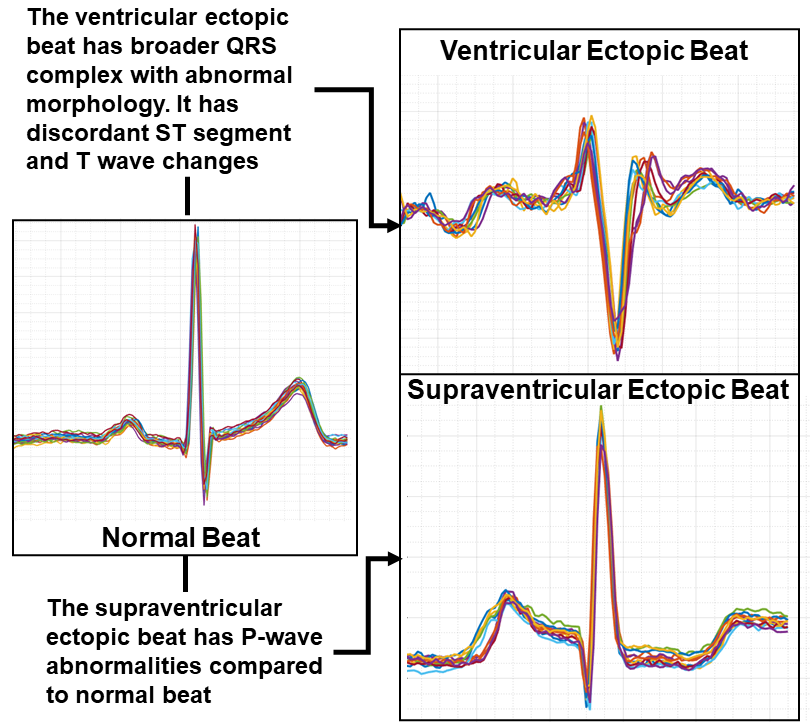}
    \caption{The waveform of normal and abnormal beats}
    \label{fig:Beat_Types}
\end{figure}

While classifying the normal and abnormal beats, we have followed the Association for the Advancement of Medical Instrumentation (AAMI) instructions, which are the golden standard for automatic heartbeats classification. According to the AAMI standard, heartbeats can be divided into N (normal), S (supraventricular ectopic beat), V (ventricular ectopic beat), F (fusion beat), and Q (unclassified beat) \cite{AAMI}. Therefore, the beats are divided into two groups for that layer. The first group only contains the normal beats N from the datasets, and the second group is composed of other types of heartbeats (S, V, F, and Q), which are the abnormal beats.
\\
If a beat is classified as normal in this stage, it is not transmitted to the fog/cloud node to save energy. Nevertheless, if a regular beat is mistakenly classified as abnormal, it can still be corrected by the next classifier. So, we need to ensure that the abnormal beats are classified with high sensitivity in that layer.  
\\
 It is known that during arrhythmias, the heart rate deviates from its normal rhythm. These heart rate variations are common in the premature beats; as the name suggests, they occur when the ventricles or atrial contract too soon, out of sequence with the normal heartbeat. Moreover, as shown in Figure \ref{fig:Beat_Types}, the waveform of the abnormal beats is different from the regular beats.
Therefore, the heart rate variability and correlation of the beats are indicators of abnormal beats. We have used these together with the result of the first output block to decide whether a beat is normal or abnormal. 
\\
If the first output block classifies a beat as abnormal, the output of the first convolutional is directly sent to the next classifier to be further examined. However, if a beat is classified as normal, other indicators just confirm the decision. The first indicator is the Heart Rate Variability (HRV) between three consecutive beats, which is calculated using Equation \ref{RR}.

\begin{equation}
     \abs{Fs \times \left(\frac{1}{RR_{(i-1)}} - \frac{1}{RR_i}\right)}
     \label{RR}
\end{equation}

\begin{figure}[h]
    \centering
    \includegraphics[scale=0.45]{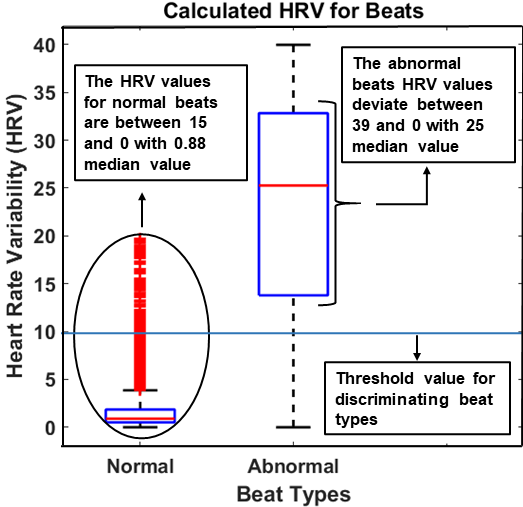}
    \caption{HRV values for normal and abnormal beats}
    \label{fig:HRV}
\end{figure}

Where $RR_i$ is the latest interval and  $Fs$ is the signal's sampling rate. Figure \ref{fig:HRV} is the whisker plot for HRV values of different beats. The threshold value is set to 10 based on variances of normal and abnormal beats. If the calculated HRV value is greater than 10, the last beat is classified as abnormal even though the first block classified it as a regular beat.  It is also observed that most of the normal beats with HRV values greater than 10 are either before or after abnormal beats.  

As a second indicator, we have calculated the correlation of a template and classified beat. While creating a template, we have used 20 different regular beats. These regular beats are aligned according to their R-peak and averaged, then the correlation is calculated using Equation \ref{Corr}.

\begin{equation}
    \rho (T,x) = \frac{1}{N-1} \sum_{n=1}^{N} \left(\frac{T[n] - \mu_T}{\sigma_T}\right) 
    \left(\frac{x[n] - \mu_x}{\sigma_x}\right)
    \label{Corr}
\end{equation}

Where $\sigma_T$ and $\mu_T$ are the standard deviation and mean of the template beat (T) and  $\sigma_x$ and $\mu_x$ are the standard deviation and mean of the incoming beat, respectively. If the correlation coefficient is lower than $0.2$, the normal beat label is changed to abnormal and forwarded to the next layer running in the fog/cloud for further classification.

\subsection{Distributed Multi-Output CNN}

\begin{figure}
    \centering
    \includegraphics[scale = 0.4]{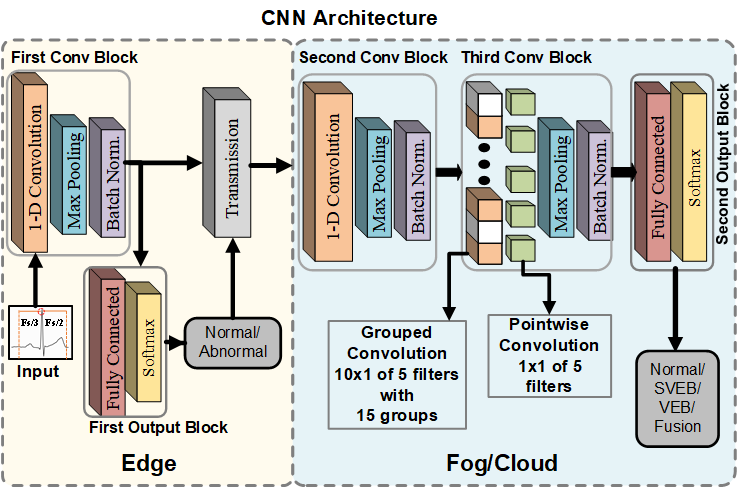}
    \caption{Distributed multi-output CNN architecture}
    \label{fig:CNN_Arch}
\end{figure}

 Our distributed multi-output CNN consists of 3 convolution and 2 output blocks. The first and last convolution blocks are followed by an output block. Figure \ref{fig:CNN_Arch} shows the layout of the designed CNN. The first and third convolution blocks consist of one convolution layer passed through ReLU activation, one max-pooling layer, and one batch normalization layer. The third convolution block is composed of depthwise and pointwise convolutions. The depthwise convolution block, which connects to each feature map, learns frequency-specific filters. The pointwise convolution block is placed after that to mix the feature maps. Also, closer inspection of the Table \ref{Tab:Multi-Output-CNN} shows that the kernel sizes of designed CNN vary from 64 to 1, which enables the model to learn different features from the heartbeats. Since the shorter filters can cover limited samples, they are used to extract more temporary and rapid oscillatory changes in the ECG, and the longer filters are used to extract more long-term features such as abnormalities in the T-wave.  Before the second output block, a dropout value of 0.2 is used to prevent overfitting during training.
Each of the output blocks consists of one fully connected layer passed through the softmax activation. The details of the architecture parameters for each of the layers are given in Table \ref{Tab:Multi-Output-CNN}.

\begin{table}[h]

	\centering
	\caption{Multi-output CNN architecture details}
	\label{Tab:Multi-Output-CNN}
	\begin{adjustbox}{width=\columnwidth,center}
	\begin{tabular}{|c|c|c|c|c|c|}
		\hline
		\textbf{Layer} & \textbf{Kernel} & \textbf{Stride} &  \textbf{Activation} & \textbf{Output} &  \textbf{\# of} \\
		\textbf{Name} & \textbf{Size} & \textbf{Size} &  \textbf{Function}& \textbf{Shape} &  \textbf{Param.} \\
		\hline\hline
			Input & - & - & - & 105x1 & 0\\
			\hline
			Conv 1 & 64 & 2 & ReLU & 53x5 & 325\\
			\hline
			Pooling 1 & 2 & 2 & - & 27x5 & 0\\
			\hline
			Batch Norm.  & - & - & - & 27x5 & 10\\
			\hline
			Fc 1 & - & - & Softmax & 2x1 & 272\\
			\hline
			\hline
			\multicolumn{5}{|c|}{\textbf{Total Number of Parameters on Edge}} & \textbf{607}\\ 
			\hline
			\hline
			Conv 2 & 32 & 1 & ReLU & 27x15 & 2415\\
			\hline
			Pooling 2 & 2 & 2 & - & 14x15 & 0\\
			\hline
			Batch Norm & - & - & - & 14x15 & 30\\
			\hline
			Grouped Conv& 10 & 1 & - &  14x75 & 825 \\
			\hline
			Pointwise Conv & 1 & 1 & Relu & 14x5 & 380 \\
			\hline
			Pooling 3 & 2 & 2 & - & 7x5 & 0\\
			\hline
			Batch Norm  & - & - & - & 7x5 & 10\\
			\hline
			Dropout & - & - & - & 7x5 & 0 \\
			\hline
			Fc 2 & - & - & Softmax & 4x1 & 144 \\
			\hline
			\hline
			\multicolumn{5}{|c|}{\textbf{Total Number of Parameters on Fog/Cloud}} & \textbf{3804}\\ 
			\hline		
		\end{tabular}
		\end{adjustbox}
	\end{table}

We design the distributed multi-output CNN considering the current limitations and resource constraints of the state-of-the-art works. It is known that the number of parameters and computational operations increases with the number of layers which leads to the consumption of more energy and memory resources. To solve these problems, we have distributed the CNN layers between edge and fog/cloud so that the introduced additional energy consumption and memory of the CNN to the edge are decreased. This fact may be seen from Table \ref{Tab:Multi-Output-CNN}. The total number of parameters before the second convolution block is 607, whereas CNN has 3804 more parameters after that. This clearly shows that by distributing the CNN over nodes, the memory requirement of the edge node can be decreased by $87\%$.

 Another advantage of this proposed solution is that we decreased the communication energy requirements by $\sim1000\times$ using the first output block. The first output block is designed to classify beats as normal or abnormal during run-time using the extracted features from the first convolution block. By adding this block, normal and abnormal beats can be distinguished without the need to invoke the other two convolutional blocks. We have observed that this classification, which uses the first convolution block, can reach up to  $95\%$ accuracy. Therefore, further processing of those regular beats would be redundant. Through avoiding this redundant operation, the system's energy consumption and inference time are decreased from both communication and computation.

\section{Experimental Setup}
\label{setup}
\subsection{Training Distributed Multi-output CNN Classifier}

We use data from MIT-BIH Arrhythmia dataset \cite{Physionet-MIT-BIH}, which contains 48 half-hours of two-channel ambulatory ECG recordings, digitized at 360 samples per second,  obtained from 47 subjects, and  MIT-BIH Supraventricular Arrhythmia Database \cite{SUPRA} which includes 78 half-hour ECG recordings with digitized at 120 Hz. We combined these two datasets to increase the number of abnormal beats. For both datasets, only the lead-II ECG signals are used for experiments. For a fair comparison with published results, we follow the evaluation settings that was most frequent in the state-of-the-art-works. We have excluded the 4 paced records (102,104,107,217) from the MIT-BIH dataset \cite{Physionet-MIT-BIH}.	
\begin{figure}[h]
    \centering
    \vspace{-3mm}\includegraphics[scale = 0.8]{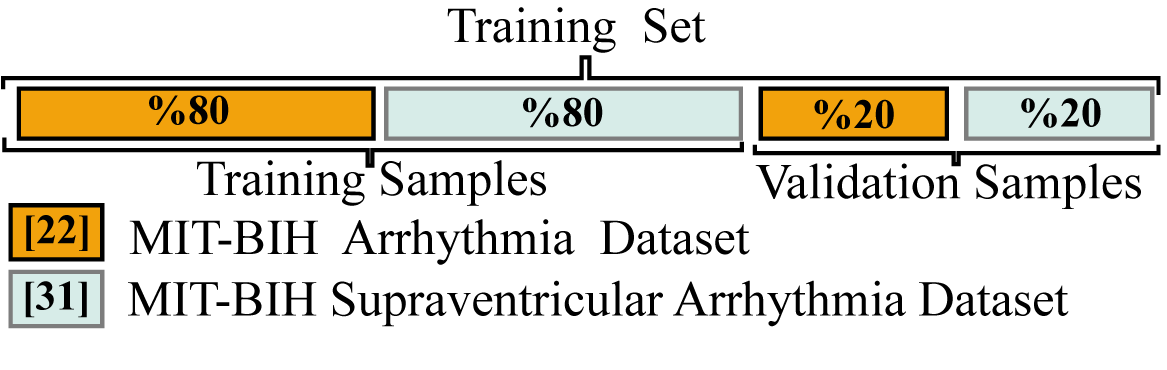}
    \caption{The ratio of training and validation samples in traning set.}
    \label{fig:setup}
\end{figure}
\vspace{-3mm}
  ECG beats in 22 recordings from the MIT-BIH dataset are included to training set. Additionally, beats from MIT-BIH Supraventricular Arrhythmia Database \cite{SUPRA} are added to the training set. In the training set, 80\% of ECG beats are used for training, and 20\% of ECG beats are used for validation (as shown in Figure \ref{fig:setup}). Since the sampling rates of the two datasets are different, we have resampled them to 130 Hz. The 44 records  (22 records from the training set and 22 records which model has never seen before) from the MIT-BIH dataset \cite{Physionet-MIT-BIH} are used as test data. The network was trained with Glorot initialization of the weights \cite{glorot}, we used the Adam optimizer \cite{Adam} with the default parameters $\beta_1=0.9$ and $\beta_2=0.999$. The learning
rate is initialized to .001 and reduced by a factor of 10 when the validation accuracy stopped improving for 15 consecutive epochs. The training continues until 100 successive epochs without validation performance improvements with a maximum of 500 epochs. The best model is chosen as the highest accuracy rate on the validation data.
After training the complete CNN architecture, the weights of the first convolution block are obtained to train the fully connected layer at the first output block. The same training data is given as input to the first convolution block again. Then, the output of the first batch normalization layer is fed to the fully connected layer instead of the second convolution block to classify beats as normal and abnormal. During backpropagation, only the weights and biases of the fully connected layer are calculated from the gradient of classification loss. In other words, while training the first output block, the parameters of the first convolutional block are not changed.

\vspace{-5mm}

\subsection{Target Wearable Device}
\label{Target Device}
Our work is designed for low-power and low-memory wearable devices. For  example,  SmartCardiaINYU \cite{Smart} device is equipped with an ultra-low-power 32-bit microcontroller STM32L151 containing an ARM Cortex–M3 with a maximum clock rate of 32 MHz. It has a 48 KB RAM, 384  KB  Flash,  and  a standard 710 mAh battery. The device captures ECG  signals using a single lead ECG sensor. We have tested our work on a \textit{BLE} standalone module similar to \cite{Smart} with a 32-bit ARM Cortex-M4 core with 40 MHz maximum operating frequency and 64 KB of RAM. Our profiling showed that algorithms peak memory usage do not exceed 30 KB of RAM.
\vspace{-3mm}
\section{Results}
\label{result}
\subsection{Performance of Noise/Artifact Detection}
Firstly, the MIT-BIH arrhythmia dataset is used to evaluate the performance. The recordings include clinically significant arrhythmias, which have quite different waveforms from the normal ECG, so it is essential to classify them as not noisy ECG signal to validate the algorithm. The performance of the proposed method is evaluated using four benchmark metrics, sensitivity (Se), accuracy (Ac), specificity (Sp), and positive predictivity (PPV) which are defined as follows:

 \begin{minipage}{.4\linewidth}
\begin{equation}
    \hspace*{-1cm} Se = \frac{TP}{TP + FN}
    \label{eq:MF1}
\end{equation}
\vspace{-6mm}
\end{minipage}%
\hspace{6mm} 
\begin{minipage}{.4\linewidth}
\begin{equation}
    Sp = \frac{TN}{TN + FP}
  \label{eq:ACC}
\end{equation}
\end{minipage}

\begin{minipage}{.30\linewidth}

\begin{equation}
    \hspace*{-1cm} PPV = \frac{TP}{TP+FP}
    \label{eq:MF1}
\end{equation}
\end{minipage}%
\hspace{0.4mm}
\begin{minipage}{.61\linewidth}
\begin{equation} 
 \hspace{0.7mm} Ac = \frac{TP + TN}{TP + TN + FN + FP}
  \label{eq:ACC}
\end{equation}
\end{minipage}

\vspace{8mm}

Where $TP$, $TN$, $FP$ and $FN$ refer to True Positive, True Negative, False Positive and False Negative, respectively.

\begin{table}[H]
\centering
\caption{Performance of the algorithm in MIT-BIH for detection of Noise/Artifact}
\begin{adjustbox}{width=\columnwidth,center}
\begin{tabular}{llllllll}
\toprule
\begin{tabular}[c]{@{}l@{}}\textbf{Segment~}\\\textbf{Type}\end{tabular} & \begin{tabular}[c]{@{}l@{}}\textbf{Total \# of}\\\textbf{Segments}\end{tabular} & \begin{tabular}[c]{@{}l@{}}\textbf{True}\\\textbf{Positives}\end{tabular} & \begin{tabular}[c]{@{}l@{}}\textbf{False}\\\textbf{Negatives}\end{tabular} & \begin{tabular}[c]{@{}l@{}}\textbf{Sp}\\\textbf{\%}\end{tabular} & \begin{tabular}[c]{@{}l@{}}\textbf{Ac}\\\textbf{\%}\end{tabular} & \begin{tabular}[c]{@{}l@{}}\textbf{Se}\\\textbf{\%}\end{tabular} &
\begin{tabular}[c]{@{}l@{}}\textbf{PPV}\\\textbf{\%}\end{tabular} \\ 
\hline
Clean  & 5000  & 4970  & 30  & 99.4  & 99.3  & 99.2 & 99.5\\
Noisy  & 5000  & 4959  & 41  & 99.2 & 99.3   & 99.4 & 99.4 \\
\hline
\end{tabular}
\end{adjustbox}
\label{MITBIHA-Noise}
\end{table}

A total of 10000 segments are obtained from the MIT-BIH dataset. As shown in Table \ref{MITBIHA-Noise}, the algorithm's accuracy can reach 99.3\% by wrongly classifying only 71 segments in the MIT-BIH dataset amongst the 10000 segments.

The PhysioNet/CinC (PICC) 2011 challenge \cite{Physionet-SQA} is used as a second dataset to validate our algorithm and compare it with the other related works. In the PICC, the ECG signals are standard 12-lead, and the leads are recorded simultaneously for 10 seconds; each lead is sampled at 500 Hz. 

\begin{table}[h]
\centering
\caption{Performance comparison of related works in PICC and MIT-BIH dataset for detection of Noise/Artifact}
\begin{adjustbox}{width=\columnwidth,center}
\begin{tabular}{c|ccc|c|c|c} 
\toprule
\textbf{Works} & \begin{tabular}[c]{@{}l@{}}\textbf{Sp}\end{tabular} & \begin{tabular}[c]{@{}l@{}}\textbf{Ac}\end{tabular} & \begin{tabular}[c]{@{}l@{}}\textbf{Se}\end{tabular} & \textbf{Dataset} & \textbf{Methods} & {\textbf{Features}} \\ 
\hline
\multirow{2}{*}{\textbf{\cite{IoT-SQA}}} & 94 & - & 99.74 & PICC & \multirow{2}{*}{Rule Based} & \multirow{2}{*}{\begin{tabular}[c]{@{}c@{}}Time \& Freq.\\Domain\end{tabular}} \\
 & 99.4 & - & 98.5 & MIT-BIH &  &  \\ 
\hline
\multirow{2}{*}{\textbf{\cite{Behar}}} & 96.5 & 97.1 & 97.7 & PICC & \multirow{2}{*}{SVM} & \multirow{2}{*}{\begin{tabular}[c]{@{}c@{}}Time \& Freq.\\Domain\end{tabular}} \\
 & 97.8 & 97.8 & 97.7 & MIT-BIH &  &  \\ 
\hline
\multirow{2}{*}{\textbf{Ours}} & 96.3 & 96 & 94.5 & PICC & \multirow{2}{*}{Rule Based} & \multirow{2}{*}{\begin{tabular}[c]{@{}c@{}}Time~\\Domain\end{tabular}} \\
 & 99.4 & 98.3 & 99.2 & MIT-BIH &  &  \\
\hline
\end{tabular}
\end{adjustbox}
\label{Tab:Noise/Artifact-Related}
\end{table}

The proposed solution to detect the noise and artifacts in the ECG signals outperforms the other related works in MIT-BIH dataset (see Table \ref{Tab:Noise/Artifact-Related}). This advantage of the algorithm is crucial since any other method that classifies arrhythmias as noisy would lead to a rejection of the data. Another advantage is that our algorithm does not require computationally heavy features to classify.

\subsection{Performance of Normal/Abnormal Beat Classification}

The performance of this layer is evaluated using the AAMI instructions for beat type classification. However, unclassified beats (Q) are excluded from the abnormal beat types since they are very rare in the dataset. As shown in Table \ref{Table:Abnormal}, this layer's classification accuracy is $98.5\%$ with a $99.6\%$ sensitivity for abnormal beats.

\begin{table}[h]
\centering
\caption{The performance evaluation of the algorithm in merged MIT-BIH for detection of Normal/Abnormal beats }
\begin{adjustbox}{width=\columnwidth,center}
\begin{tabular}{lllllll|l} 
\toprule
\multicolumn{1}{l}{\textbf{Labels}} & \begin{tabular}[c]{@{}l@{}}\textbf{Total \#}\\\textbf{Beats}\end{tabular} & \begin{tabular}[c]{@{}l@{}}\textbf{True}\\\textbf{Classified}\end{tabular} & \begin{tabular}[c]{@{}l@{}}\textbf{False}\\\textbf{Classified}\end{tabular} & \begin{tabular}[c]{@{}l@{}}\textbf{Beat}\\\textbf{Type}\end{tabular} & \begin{tabular}[c]{@{}l@{}}\textbf{Se}\\\textbf{\%}\end{tabular} & \multicolumn{1}{l}{\begin{tabular}[c]{@{}l@{}}\textbf{Sp}\\\textbf{\%}\end{tabular}} & \begin{tabular}[c]{@{}l@{}}\textbf{Acc}\\\textbf{\%}\end{tabular} \\ 
\toprule
Abnormal & 18000 & 17929 & 71 & S,V,F & 99.6 & 97 & \multirow{2}{*}{98.5} \\
Normal & 15000 & 14545 & 455 & N & 97 & 99.6 &  \\
\bottomrule
\end{tabular}
\end{adjustbox}
\label{Table:Abnormal}
\end{table}

The goal in this layer is to achieve high sensitivity for abnormal beats classification because if a beat is classified as normal, it would not be transmitted to the fog/cloud node. Table \ref{Table:Abnormal} shows that amongst the 18000 abnormal beats, only 71 of them are misclassified, which indicates a very high sensitivity for the abnormal beat classification.

\begin{table}[h]
\centering
\caption{The performance comparison of related works with ours for Abnormal beat detection}
\begin{tabular}{clccc} 
\toprule
\multicolumn{1}{l}{\textbf{Works}} & \textbf{Methods} & \begin{tabular}[c]{@{}l@{}}\textbf{Se}\end{tabular} & \begin{tabular}[c]{@{}l@{}}\textbf{PPV}\end{tabular} & \begin{tabular}[c]{@{}l@{}}\textbf{Acc}\end{tabular} \\ 
\toprule
\textbf{\cite{Remote-2}} & SVM & - & - & 96 \\ 
\hline
\textbf{\cite{Abnormal-2}} & \begin{tabular}[c]{@{}l@{}}Tree\end{tabular} & 96.4 & 92 & - \\ 
\hline
\textbf{Ours} & \begin{tabular}[c]{@{}l@{}}CNN \& \\Rule based\end{tabular} & 97 & 99.6 & 98.5 \\
\hline
\end{tabular}
\label{Tab:RelatedWork-Abnormal}
\end{table}

Table \ref{Tab:RelatedWork-Abnormal} compares the performance of related works with the proposed solution on the MIT-BIH dataset. The proposed algorithm outperforms others in terms of classification performance. Moreover, these methods use feature-based classifiers such as SVM and tree, so they need to extract different features from the ECG signal. For example, the authors in \cite{Remote-2} extract 14 different time-frequency domain features from ECG, requiring additional memory and a computational overhead which does not suit the edge device. We tried to evaluate the memory  and energy consumption of the state-of-the-art works in our target wearable device, which is explained in Section \ref{Target Device}. However, the target device's memory overflowed more than $1.5\times$ of its maximum memory for all algorithms \cite{Remote-2, Abnormal-2} due to huge memory requirements for the feature extraction and classification. 
On the other hand, our proposed solution calculates the correlation between template and input beats, checks the heart rate variability; no additional features are extracted from a beat specific to that layer. Moreover, unlike the related works, we do not use a feature-based classifier since the first convolution block is used to classify the beat types (N, S, V, F) for complete CNN architecture. The designed first convolution and output blocks consist of 607 parameters that allow us to run that algorithm on edge. 
\vspace{-3mm}
\subsection{Performance of CNN Classifier}
\begin{figure}[h]
    \centering
    \includegraphics[scale = 0.47]{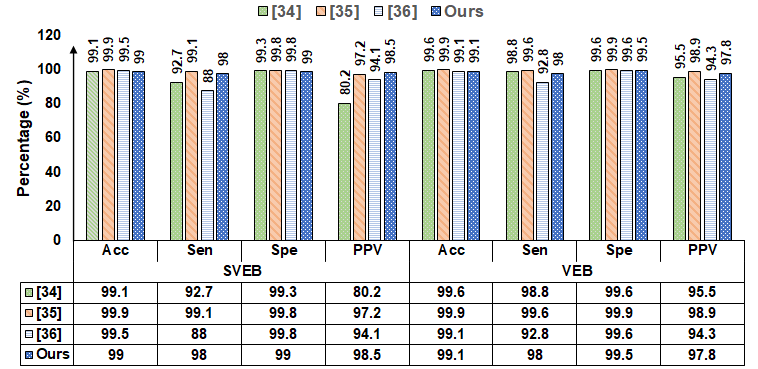}
    \caption{Performance comparison between the proposed methodology and related works}
    \label{fig:Performance-CNN}
\end{figure}

The overall performance is evaluated according to the second output block, which classifies beats into Normal, SVEB (supraventricular ectopic heartbeats), VEB (ventricular ectopic heartbeats), and Fusion beat. To make a fair comparison with other works, the performance of $Normal/Abnormal$ $Beat$ $Classifier$ is also considered. So, if a beat is misclassified as normal in that layer, it is not transmitted to further CNN layers. The classification performance (Acc, Sen, Spe, Ppr) of VEB and SVEB are also investigated to be consistent with related works and given in Figure  \ref{fig:Performance-CNN}.

% \begin{table}[H]
% \centering
% \caption{Performance comparison between the proposed methodology and related works}
% \begin{tabular}{c|cccc|llll} 
% \toprule
% \multirow{2}{*}{\textbf{Works}} & \multicolumn{4}{c|}{\textbf{SVEB}} & \multicolumn{4}{c}{\textbf{VEB}} \\
%  & Acc & Sen & Spe & PPV & Acc & Sen & Spe & PPV \\ 
% \toprule
% \textbf{\cite{EMBC}} & 99.1 & 92.7 & 99.3 & 80.2 & 99.6 & 98.8 & 99.6 & 98.9 \\
% \textbf{\cite{Access}} & 99.3 & 99.1 & 99.8 & 97.2 & 99.9 & 99.6 & 99.9 & 98.9 \\
% \textbf{\cite{BCS}} & 99.5 & 88 & 99.8 & 94.1 & 99.1 & 92.8 & 99.6 & 94.3 \\
% \textbf{Proposed} & 99 & 98 & 99 & 98.5 & 99.1 & 98 & 99.5 & 97.8 \\
% \hline
% \end{tabular}
% \label{Tab:Performance-CNN}
% \end{table}

\begin{figure}[h]
    \centering
    \includegraphics[scale=0.5]{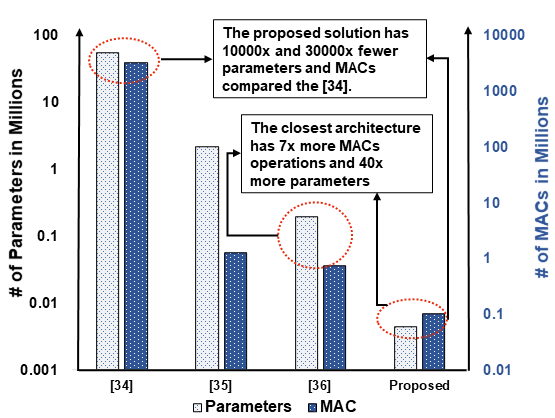}
    \caption{Comparison of computational complexity and memory with related works }
    \label{fig:MACs}
\end{figure}

Figure \ref{fig:Performance-CNN} shows that the proposed algorithm's performance is the highest for \textit{PPV} of SVEB; and for the other metrics, it is $0.4\% - 1\%$ below from the best one. Also, closer inspection of the table shows that the performance of \cite{Access} reaches more than $0.999$ for 3 different metrics which is practically hard to outperform. However, our algorithm's worst performance is $97.8\%$ whereas it is $80.2\%$ and $88\%$ for \cite{EMBC} and \cite{BCS}, respectively. Since the number of normal beats in MIT-BIH is approximately 10 times that of abnormal data, it is easy for models to achieve high accuracy. However, it is harder to achieve high performance on all the metrics. Also, the other models, in general, are not suitable for the edge devices as their architecture has many parameters. To compare models in terms of memory and energy, we estimated the number of parameters and multiply-accumulate operations (MACs) of each architecture. Equation \ref{MAC} is used for the calculation of MACs,

\vspace{-5mm}
\begin{equation}
    MACs = \frac{C_{in} \times C_{out} \times K_h \times K_w \times H_{out} \times W_{out} }{g}
    \label{MAC}
\end{equation}
\vspace{-4mm}

where $C_{in}$ is the number of input channels, $C_{out}$ is the number of output channels, $H_{out}$ and $W_{out}$ are the height and width of the layer’s output, respectively. $K_h$ × $K_w$ is the kernel size of
each convolution, and $g$ is the number of groups if there are any.
When the MACs are calculated for different works, we observed that the proposed solution has $30000\times$ fewer operations compared to \cite{EMBC}. Therefore, for better visualization, Figure \ref{fig:MACs} is given in logarithmic form where the two y-axes are in millions.
Figure \ref{fig:MACs} shows that the proposed solution is much more energy and memory efficient compared to other existing work. Even the closest CNN architecture performs $7\times$ more $MAC$ operations and requires $40\times$ more memory compared to ours.
\\
The proposed CNN has very few parameters and MACs compared to related works because of two reasons. First, in the third convolution block of the proposed CNN, the grouped and pointwise convolutions are performed instead of 1D convolutions to mix the high-level features optimally. These convolution operations are also more efficient compared to the 1D convolutions since the number of $MACs$ are decreased with increasing groups in convolutions. Second, as we applied a bandpass filter with cut-off frequencies $f_1=1$ Hz and $f_2=50$ Hz at the beginning of the process, the ECG signal's sampling rate, which directly affects the input size of CNN, is downsampled to 130 Hz without losing any information.
\vspace{-4mm}
\subsection{Memory and Energy Consumption Evaluation}

We evaluate the memory footprint and energy consumption of our proposed methodology on the  target device mentioned in Section \ref{Target Device}. Table \ref{Tab:Hardware} shows the execution time, energy consumption, and required memory for each layer that runs on the edge device. When we evaluate the power and execution time of each layer, we perform multiple experiments to take the average of them. In the end, we observe $\pm 1\%$ mW and $\pm 0.5\%$ ms deviation from the average of all trials. For example, we observe a maximum $20.12$ ms execution time and $14.13$ mW average power consumption for the First Convolutional Block. The model is implemented and deployed to the target device using MATLAB (MATLAB and Coder Toolbox Release R2020b,The MathWorks, Inc, USA).

% \begin{adjustbox}{width=\columnwidth,center}
% $\geq$ 8 KB

\begin{table}[H]
\centering
\caption{Memory and energy consumption on blue Gecko}
\begin{adjustbox}{width=\columnwidth,center}
\begin{tabular}{|c|c|c|c|c|c|} 
\hline
\textbf{Layers} & \textbf{Operation} & \begin{tabular}[c]{@{}c@{}}\textbf{Exe.}\\\textbf{Time (ms)}\end{tabular} & \begin{tabular}[c]{@{}c@{}}\textbf{Average}\\\textbf{Power (mW)}\end{tabular} & \begin{tabular}[c]{@{}c@{}}\textbf{Energy}\\\textbf{($\mu$J)}\end{tabular} & \begin{tabular}[c]{@{}c@{}}\textbf{Compatible~}\\\textbf{RAM}\end{tabular} \\ 
\hhline{|======|}
\begin{tabular}[c]{@{}c@{}}\textbf{Noise/Motion}\\\textbf{Artifact }\end{tabular} & \begin{tabular}[c]{@{}c@{}}Feature\\Extraction\end{tabular} & 13.5 & 14.3 & 193.05 & $\geq$ 32 KB \\ 
\hline
\begin{tabular}[c]{@{}c@{}}\textbf{ Normal/}\\\textbf{Abnormal beat}\end{tabular} & \begin{tabular}[c]{@{}c@{}}Template Check\\\& HRV\end{tabular} & 1.8~ & 23 & 41.4 & $\geq$ 8 KB \\ 
\hline
\multirow{2}{*}{\vspace{-3mm}\textbf{CNN }} & \begin{tabular}[c]{@{}c@{}}First Convolution\\Block\end{tabular} & 20 & 14.1 & 282 & \multirow{2}{*}{\vspace{-3mm}$\geq$ 32 KB} \\ 
\cline{2-5}
 & \begin{tabular}[c]{@{}c@{}}First Output\\~Block\end{tabular} & 1.2 & 7.5 & 9 &\\
\hline
\end{tabular}
\end{adjustbox}
\label{Tab:Hardware}
\end{table}

The overall execution time for a heartbeat takes 36 ms in the edge device with 55 mW power consumption. Also, our proposed methodology is compatible with any devices with a minimum RAM of 32 KB. As a result, our methodology guarantees high classification performance while maintaining the low-power wearable devices requirements of being resource-efficient in terms of energy and memory.
\vspace{-3.5mm}
\section{Discussion and Future Work}
In this paper, we present a novel and energy-efficient methodology that runs on a hybrid edge-fog/cloud architecture for continuously monitoring the heart at low-power wearable devices. To evaluate our methodology's performance, we compare our approach with several state-of-the-art methods that evaluate their classification results on the same datasets. We show that our proposed methodology reaches or outperforms the current state-of-the-art works in terms of classification performance for 3 different tasks (Noise/Artifact detection, Normal/Abnormal beat detection, Abnormal beat classification) while being energy and memory efficient.
However,  despite these promising results, questions remain about whether the proposed approach's performance is excellent. Therefore, it is important to evaluate the limitations of our methodology. 
\\[3pt]
Firstly, in our proposed methodology, the Abnormal beat detection and classification layers heavily depend on the R-peak detection performance. It is observed that when the detected R-peaks are wrong, the classification performance decreases severely due to wrong segmentation and HRV calculation. Secondly, even though the MIT-BIH dataset is widely used in literature, most state-of-the-art-works and our proposed methodology focus on identifying small numbers of cardiac abnormalities (VEB, SVEB) that do not represent the complexity and difficulty of ECG interpretation. Therefore, we believe that there is abundant room for further progress in beat classification in wearable devices. For example, this paper showed that the HRV is a helpful feature to classify beats as normal or abnormal. In future investigations, it might be helpful to use different machine learning structures such as neural graph learning to integrate heart rate variability features into an end-to-end model. Since our proposed methodology is a distributed neural architecture between nodes of the network, a further study with more focus on federated learning can be employed to increase performance while preserving privacy \cite{fede}. Also, the performance of the transmission decision unit (Normal/Abnormal beat classification) can be further studied using a more comprehensive ECG dataset with different arrhythmias and abnormalities.

\vspace{-3mm}
\section{Conclusion}
\label{conclusion}
This paper proposes a methodology for real-time continuous heart monitoring using distributed multi-output CNN. The neural network layers are distributed between edge-fog/cloud so as to save energy and time while reducing the communication channel usage and server load. Moreover, the proposed methodology requires $40\times$ less memory compared to state-of-the-art works while maintaining high accuracy.  To the best of our knowledge, our methodology achieves the best performance on heartbeat classification while being $7\times$ more energy efficient for devices with a minimum of 32 KB of RAM.

\end{document}